\documentclass[12pt]{article}
\usepackage{amsmath}
\usepackage{amsfonts}
\usepackage{amsthm}
\begin{document}
\def\Tm{T^m}
\def\tm{t_m}
\def\Tdm{T^m_d}
\def\tdm{t_m^d}
\def\ot{\otimes}
\def\br{\mathbb{R}}
\def\al{\alpha}
\def\bt{\beta}
\def\th{\theta}
\def\ga{\gamma}
\def\vth{\vartheta}
\def\de{\delta}
\def\lm{\lambda}
\def\b{\beta}
 \def\tp{{\rm tg}({\phi\over 2})}
\def\k{\kappa}
 \def\pp{{\pi\over 2}}
\def\om{\omega}
\def\si{\sigma}
\def\w{\wedge}
\def\od{\sqrt{2}}
\def\Tn{T^n}
\def\tn{t_n}
\def\Tdn{T^n_d}
\def\tdn{t_n^d}
\def\e{\varepsilon}
\def\ti{\tilde}
\def\js{{1\over 4}}
 \def\D{{\cal D}}
\def\I{{\cal I}}
\def\L{{\cal L}}
\def\ri{{\mathrm{i}}}      
\def\S{{\cal S}}
 \def\H{{\cal H}}
\def\G{{\cal G}}
\def\bz{{\bar z}}
\def\E{{\cal E}}
\def\B{{\cal B}}
\def\M{{\cal M}}
 \def\A{{\cal A}}
\def\K{{\cal K}}
\def\J{{\cal J}}
\def\tJ{\ti{\cal J}}
\def\R{{\cal R}}
\def\d{\partial}
\def\la{\langle}
\def\ra{\rangle}
\def\bc{{\mathbb C}}
\def\st{\stackrel{\w}{,}}
\def\lta{\leftarrow}
\def\rta{\rightarrow}
\def\scu{$SL(2,\bc)/SU(2)$~ $WZW$  }
\def\xpm{\partial_{\pm}}
\def\xp{\partial_+}
 \def\xm{\partial_-}
 \def\ps{\partial_{\sigma}}
  \def\pt{\partial_{\tau}}
\def\be{\begin{equation}}
\def\ee{\end{equation}}
\def\jp{\frac{1}{ 2}}
\def\noi{\noindent}
\def\nl{\nabla}
 
\def\tD{\Delta^*}
\def\slc{SL(2,{\bf C})}
\begin{titlepage}
\begin{flushright}
{}~
  
\end{flushright}

\vspace{1cm}
\begin{center}
{\Huge \bf Integrability of the bi-Yang-Baxter $\sigma$-model}\\ 
[50pt]{\small
{ \bf Ctirad Klim\v{c}\'{\i}k}
\\
Aix Marseille Universit\'e, CNRS, Centrale Marseille\\ I2M, UMR 7373\\ 13453 Marseille, France}
\end{center}

\vspace{0.5 cm}
\centerline{\bf Abstract}
\vspace{0.5 cm}
\noindent    We construct a Lax pair with spectral parameter for  a  two-parameter
doubly Poisson-Lie   deformation of the principal chiral model.

\end{titlepage}
 \section{Introduction}
 
Let $\G$ be a simple compact Lie algebra  with its Killing-Cartan form $(.,.)_\G$. An $\br$-linear  map $R:\G\to\G$   is called a  Yang-Baxter  operator if it satisfies
 the skew-symmetry condition
  \be (RX,Y)_\G+(X,RY)_\G=0\label{skew}\ee
  and  the  following variant of the modified Yang-Baxter equation
   \be [RX,RY]=R([RX,Y]+[X,RY])+[X,Y], \quad X,Y\in \G. \label{myb}\ee
   The canonical example of the Yang-Baxter operator is given by 
    \be   RT^\mu=0,\quad RB^\al=C^\al,\quad RC^\al=-B^\al,\label{R}\ee
   where 
   \be  T^\mu=\ri H^\mu,\quad B^{\al}=\frac{\ri}{ \od}(E^{\al}+E^{-\al}),\quad 
C^{\al}=\frac{1}{ \od}(E^{\al}-E^{-\al}),\ee
 and   $(H^\mu,E^\al)$ is 
 the usual Cartan-Weyl basis of the complex Lie algebra 
$\G^{\mathbb C}$ (see Sec. 2.2. of \cite{K09} for more details).

   Yang-Baxter operators   play important role in the theory of integrable systems (cf.  e.g.  \cite{RST}), but it was only recently when they entered in the
   description of {\it geometries} of target spaces of certain integrable non-linear $\sigma$-models in $1+1$-dimensions \cite{K02,K09}.    The integrable models in question were baptised the Yang-Baxter   $\sigma$-models accordingly and, for a given simple compact group $G$,  the action of such a model reads \cite{K02}
  \be S_\beta(g)= \int_W (g^{-1}\xp g,(I-\beta R)^{-1} g^{-1}\xm g)_\G.\label{YB}\ee
 Here  $I:\G\to\G$ is the identity map, $\partial_\pm$ are the   derivatives   with respect to the light-cone coordinates $\xi_\pm$ on the  world-sheet $W$ and $g(\xi_+,\xi_-)$ is a group-valued field
 configuration. We observe that for $\beta=0$ we recover from \eqref{YB} the action of the principal chiral model \cite{ZM}.
 
 The Yang-Baxter $\sigma$-model \eqref{YB} was  subsequently reobtained by Delduc, Magro and Vicedo  \cite{DMV13} in a new way, which made more  transparent its integrability and also its symmetry structure. Indeed,  though  the $\beta$-deformation breaks the   right $G$-symmetry of the principal chiral model   (while the left symmetry is kept intact), the right translations still  continue to   act as the so called  Poisson-Lie symmetries \cite{K09}. The corresponding conserved Poisson-Lie  charges were then explicitely computed in \cite{DMV13}.  Delduc, Magro et Vicedo
 have also generalized the Yang-Baxter deformation  to the coset target spaces  and even to the  supercoset target $AdS_5\times S^5$ important for the superstring theory \cite{DMV13b}.  Further
 superstring applications of this result followed readily \cite{ABF,KMY14}. 
 
 Is there a two-parameter deformation of the principal chiral model which would convert both left and right symmetries into the Poisson-Lie symmetries? The answer to this question is affirmative and
 the corresponding "bi-Yang-Baxter $\sigma$-model" was constructed in \cite{K09}. Its action reads
  \be S_{\alpha,\beta}(g)= \int_W (g^{-1}\xp g,(I-\alpha R_g-\beta R)^{-1} g^{-1}\xm g)_\G,\label{BYB}\ee
 where $R_g\equiv$ Ad$_{g^{-1}}R$Ad$_g$. 
 
 In this article, we  establish the  integrability of the bi-Yang-Baxter $\sigma$-model  making it the  first  known integrable model on a group  manifold having
  no apparent  non-Abelian symmetries.  We note in this respect, that the very few integrable $\sigma$-models on   group targets  known   up to now  are  all symmetric with respect to an appropriate action of the group on which they live. This is the case 
 for  specific metric deformations of principal chiral model on the $SU(2)$ group \cite{Ch,KY,KMY11} and on the
 Nappi-Witten group \cite{Moh}, respectively,  for the Yang-Baxter $\sigma$-model  on every simple compact group \cite{K02,K09} and, finally, for the model interpolating between
 exact WZNW CFT and non-Abelian T-dual of principal chiral model  on every simple compact group \cite{BFHP, EH, S}. 
 
 The lack of symmetry of the bi-Yang-Baxter $\sigma$-model   makes  difficult to  search a suitable Lax pair  starting from some ansatz,   since it is not obvious how  such ansatz should depend on the group element $g$.  For this precise reason we left the problem open in \cite{K09} and decided to re-examine it only when   the paper \cite{DMV13}  appeared.  We hoped to adapt the reasoning of Delduc, Magro and Vicedo to the  presence of two deforming paramaters $\al$ and $\beta$ but, quite unexpectedly, our reimmersion in the old stuff     permitted us to identify the needed Lax pair by further developing the methods used   already in \cite{K09}!   We shall thus  argue,  that the  bi-Yang-Baxter Lax pair has    the following form
   \be L^{\al,\bt}_\pm(\zeta) =\mp \biggl(\beta(R-\ri)+ \frac{2\ri \bt \pm(1+\al^2-\bt^2)}{1\pm\zeta}\biggr) (I\pm\al R_g\pm \bt R)^{-1}g^{-1}\partial_{\pm}g,\label{dLax}\ee
   where $\zeta$ is a complex valued spectral parameter. We note that for $\al=\bt=0$ the bi-Yang-Baxter $\si$-model (\ref{BYB}) becomes the principal chiral model and 
 (\ref{dLax}) becomes the standard Lax pair introduced by Zakharov and  Mikhailov in \cite{ZM}:
 \be  L^0_\pm(\zeta) =-\frac{g^{-1}\xpm g}{1\pm \zeta}.\label{ZM}\ee

The plan of the article is as follows: in Section 2, we first review   the dynamics of the bi-Yang-Baxter $\sigma$-model, write  its field equations and its Bianchi identities in terms of suitable currents
and prove that the expression \eqref{dLax}  gives  indeed the bi-Yang-Baxter  Lax pair.  We then discuss some limiting cases of the parameters $\al$ and $\bt$ and, as a  by-product of this discussion,  we  establish the gauge  equivalence of  the Lax pairs  of the  simple Yang-Baxter $\sigma$-model  obtained in \cite{K09} and in \cite{DMV13}, respectively. In Section 3,    
we discuss the notion of the extended solution of the Yang-Baxter $\sigma$-model and show how the bi-Yang-Baxter Lax pair can be actually {\it derived} from it, so that no guess-work was necessary to obtain the quite complicated expression \eqref{dLax}. 
  We  finish by  a short  outlook.

\section{ Lax pair of the bi-Yang-Baxter $\sigma$-model}

 We begin by noting that the relations \eqref{skew} and \eqref{myb} imply
 that  the antisymmetric bracket 
  \be [X,Y]_R\equiv [RX,Y]+[X,RY], \quad X,Y\in \G\ee 
 verifies the Jacobi identity  and, hence, it defines a new Lie algebra structure  $\G_R\equiv (\G,[.,.]_R)$ on the
  vector space $\G$.  We know introduce $\G$-valued "currents" by the prescription:
  \be J_\pm:=  \mp(I\pm\al R_g\pm \bt R)^{-1}g^{-1}\partial_{\pm}g.\label{cur}\ee
  By varying the bi-Yang-Baxter action, we find that the corresponding field equations can be written as the following $\G_R$ zero curvature condition:
 \be  \partial_+J_--\partial_-J_++\bt[J_-,J_+]_R=0.\label{FE}\ee
 It follows from the definition \eqref{cur} of the currents  $J_\pm$ that they fulfil certain Bianchi identities.  They are given by:

\medskip

 \noindent {\bf Lemma}:  {\it For every solution $g(\xi_+,\xi_-)$ of the field equations \eqref{cur},\eqref{FE} of the   bi-Yang-Baxter model \eqref{BYB}, the currents \eqref{cur} verify also the following  Bianchi identity:}
 \be \d_+J_-+\d_-J_+  +\bt [J_-,RJ_+] +\bt [J_+,RJ_-] +(1+\al^2-\bt^2)[J_-,J_+]=0.\label{BI}\ee

 \begin{proof}
 
 We start with the obvious Maurer-Cartan  identity:
\be  \partial_+(g^{-1}\partial_-g)-\d_-(g^{-1}\d_+g)-[g^{-1}\partial_-g,g^{-1}\partial_+g]=0\ee
which, using the definition \eqref{cur}, can be rewritten  as follows
$$\d_+((1-\al R_g-\bt R)J_-) +\d_-((1+\al R_g+\bt R)J_+) +$$\be + [(1-\al R_g-\bt R)J_-,(1+\al R_g+\bt R)J_+]=0.\label{ime}\ee
Using again  \eqref{cur}, we  now   evaluate:
$$ \d_+(R_gJ_-)=$$$$=\d_+(g^{-1}(R(gJ_-g^{-1}))g)=-[g^{-1}\d_+g, R_gJ_-]+R_g(\d_+J_-) +R_g[g^{-1}\d_+g,J_-]=$$
 \be =[(1+\al R_g+\bt R)J_+,R_gJ_-]+R_g(\d_+J_-)-R_g[(1+\al R_g+\bt R)J_+,J_-].\label{ui+}\ee
 Similarly, we find
 $$ \d_-(R_gJ_+)=$$
 \be  =-[(1-\al R_g-\bt R)J_-,R_gJ_+]+R_g(\d_-J_+)+R_g[(1-\al R_g-\bt R)J_-,J_+]\label{ui-}.\ee
 Inserting \eqref{ui+} and \eqref{ui-} into \eqref{ime}, we  obtain
 $$\d_+J_-+\d_-J_++ [J_-,J_+]  +\bt [J_-,RJ_+] +\bt [J_+,RJ_-]  $$
 $$ -(\al R_g +\bt R) (\d_+J_--\d_-J_+ + \bt [J_-,J_+]_R) +$$
 $$+\bt^2 \biggl([RJ_+, RJ_-]-R[RJ_+,J_-]-R[J_+,RJ_-]\biggr)$$
 \be  -\al^2\biggl([R_gJ_+,R_gJ_-]-R_g[R_gJ_+,J_-]-R_g[J_+,R_gJ_-]\biggr)=0.\label{HE}\ee
 The second line of \eqref{HE} vanishes because of  the field equations \eqref{FE} and the last two lines simplify because of the modified Yang-Baxter equation \eqref{myb}
 (the modified Yang-Baxter equation is obviously verified also by the operator $R_g$). We thus recover from \eqref{HE} the desired Bianchi identity 
  \eqref{BI}.
  \end{proof}
  
  \medskip
  
  In what follows, it will be convenient to introduce two current dependent expressions $V_\pm$:
  \be  V_\pm:= \pm \d_\pm J_\mp\pm \bt[J_\mp,RJ_\pm] \pm\jp(1+\al^2-\bt^2)[J_-,J_+].\ee
  Note that the field equations \eqref{FE} and the Bianchi identities \eqref{BI} can be cast, respectively, as 
  \be V_++V_-=0,\qquad V_+-V_-=0.\label{VV}\ee

  \noindent {\bf Theorem (Lax pair)}: 
 {\it   The following expressions give  the Lax pair of the bi-Yang-Baxter  $\sigma$-model:
  \be L^{\al,\bt}_\pm(\zeta) =\mp \biggl(\beta(R-\ri)+ \frac{2\ri \bt \pm(1+\al^2-\bt^2)}{1\pm\zeta}\biggr) (I\pm\al R_g\pm \bt R)^{-1}g^{-1}\partial_{\pm}g.\label{dLaxb}\ee
This means, in other words, that 
for every solution $g(\xi_+,\xi_-)$ of the field equations of the   bi-Yang-Baxter model \eqref{BYB} and for a  generic value of the complex  spectral parameter $\zeta$
the  following zero-curvature condition holds:}
  \be \d_+L_-^{\al,\bt}(\zeta)-\d_-L_+^{\al,\bt}(\zeta)+  [L_-^{\al,\bt}(\zeta),L_+^{\al,\bt}(\zeta)]=0.\ee
   \begin{proof}
 Using  \eqref{cur}, we first rewrite 
  \be L^{\al,\bt}_\pm(\zeta) =  \biggl(\beta(R-\ri)+ \frac{2\ri \bt \pm(1+\al^2-\bt^2)}{1\pm\zeta}\biggr) J_\pm\label{dLaxc}\ee
  and then we use the modified Yang-Baxter equation \eqref{myb} to calculate
 straightforwardly:
  $$\d_+L_-^{\al,\bt}(\zeta)-\d_-L_+^{\al,\bt}(\zeta)+  [L_-^{\al,\bt}(\zeta),L_+^{\al,\bt}(\zeta)]=$$
  \be=\biggl(\beta(R-\ri)+ \frac{2\ri \bt +(1+\al^2-\bt^2)}{1+\zeta}\biggr)V_-+\biggl(\beta(R-\ri)+ \frac{2\ri \bt -(1+\al^2-\bt^2)}{1-\zeta}\biggr)V_+.\ee
  Eqs. \eqref{VV} then imply
 that both  $V_\pm$ vanish  when $g(\xi_+,\xi_-)$ is a solution of the field equations of the   bi-Yang-Baxter model \eqref{BYB}.
  \end{proof}
  
  Let us now discuss the integrability of the  bi-Yang-Baxter  models for some special values of the parameters $\al$ and $\bt$. 
   We have already mentioned in the Introduction that  for $\al=\bt=0$  the Lax pair \eqref{dLaxb}  becomes the Zakharov-Mikhailov Lax pair  \eqref{ZM}.   
  For $\bt=0$ the action \eqref{BYB}  of the bi-Yang-Baxter model becomes 
    \be S_{\alpha}(g)= \int_W (g^{-1}\xp g,(I-\alpha R_g)^{-1} g^{-1}\xm g)_\G.\label{YBa}\ee
    This does not look like the Yang-Baxter $\sigma$-model \eqref{YB}, but,  in fact, it coincides with it upon the transformation $g\to g^{-1}$ and the change of the parameters $\al\to\bt$.
    The action of the Yang-Baxter $\sigma$-model was  written in the form \eqref{YBa} in \cite {D} where also the Lax pair of this model was written as
      \be L^{\al}_\pm(\zeta) =-    \frac{1+\al^2}{1\pm\zeta} (I\pm\al R_g)^{-1}g^{-1}\partial_{\pm}g.\label{dLaxD}\ee
      We thus observe with satisfaction  that our bi-Yang-Baxter Lax pair \eqref{dLaxb} gives for $\bt=0$  precisely  the expression \eqref{dLaxD}.
      
      In the limit  $\al=0$, we obtain in turn 
      \be L^{\bt}_\pm(\zeta) = \biggl(\beta^2 \mp \bt R -\frac{1+\bt^2}{1\pm \lm(\zeta)}\biggr) (I \pm \bt R)^{-1}g^{-1}\partial_{\pm}g,\label{dLaxJ}\ee
      where 
      \be \lm(\zeta)= \frac{\zeta-\ri \bt}{1-\ri \zeta \bt}\ .\label{pro}\ee
    We observe, that  Eq. \eqref{dLaxJ}  gives nothing but the Lax pair of the Yang-Baxter $\sigma$-model \eqref{YB} as given in \cite{K09}.  
    
    Why two differently looking expressions \eqref{dLaxD} and \eqref{dLaxJ}  can be both Lax pairs of the same model? This happens because the Lax pair of an integrable model   is  not  defined uniquely.  Apart from the
    linear fractional transformations of the spectral parameter of the type \eqref{pro}, also any gauge tranformation   of the Lax connection  gives a good Lax pair. Indeed, let us calculate
    the gauge transformation of \eqref{dLaxD}, which obviously preserves the zero-curvature condition:
          \be L^{\al}_\pm(\zeta) \to gL^{\al}_\pm(\zeta)g^{-1} +\d_\pm gg^{-1} =- \biggl(\al^2 \mp \al R -\frac{1+\al^2}{1\pm \zeta^{-1}}\biggr) (I \pm \al R)^{-1}\partial_{\pm}gg^{-1}.
      \label{inn}   \ee
     The transformation $g\to g^{-1}$, $\al\to\bt$   that transforms  the  version \eqref{YBa}  of the  Yang-Baxter $\sigma$-model  into the version \eqref{YB},  transforms also  (up to the inessential linear fractional transformation of the spectral parameter)   the Lax pair \eqref{inn} into  the Lax pair 
     \eqref{dLaxJ}.  In this way we have proven the equivalence of the Lax pairs
      of the Yang-Baxter $\sigma$-model obtained in \cite{K09} and in \cite{DMV13,D}.
         
       Finally, we comment the last special case $\al=\bt$. While the one-parameter  Yang-Baxter deformation breaks the left-right symmetry of the principal chiral model (i.e. the symmetry with respect to the  field transformation $g\to g^{-1}$), the bi-Yang-Baxter model for $\al=\bt$ does preserve this discrete symmetry. We expect that  this fact will bring about new special properties
       of the model for $\al=\bt$.
       
       \section{Extended solutions}
       The present article would not be complete, if we did not explain, how we have found the bi-Yang-Baxter  Lax pair  \eqref{dLaxb}. In fact, it was  by no means  the fruit of a guess work
       but rather  the exploitation of properties of the so-called extended solutions of the integrable $\sigma$-models. The concept of the extended solution \cite{U} plays 
 an important role  in the studies of the principal chiral model, in particular in connection with the
 so called dressing symmetries \cite{ZM,ST,Ma, DS, SV}.   We  shall first explain what is the extended solution of the principal chiral model and how to obtain the Lax pair of the
 Yang-Baxter $\sigma$-model out of it and    then we shall consider  the extended solution
 of the Yang-Baxter $\sigma$-model and obtain the Lax pair of the bi-Yang-Baxter $\sigma$-model out of it.

 Let  $g_0:W\to G$ be an ordinary  solution  of the  principal chiral model and consider the associated 
  Zakharov-Mikhailov Lax pair (\ref{ZM}):
   \be  L^0_\pm(\zeta) =-\frac{g^{-1}\xpm g}{ 1\pm \zeta}.\label{ZMb}\ee 
   Because $g_0$ is solution, the zero curvature  condition holds for generic $\zeta$:
    \be \d_+L_-^0(\zeta)-\d_-L_+^0(\zeta)+  [L_-^0(\zeta),L_+^0(\zeta)]=0 \label{cco}\ee
    hence there exists a map $l_0(\zeta)$ from   the (simply connected)  world-sheet $W$
  to the complexified group $G^\bc$ such that
 \be - l_0^{-1} (\zeta)\xpm l_0 (\zeta) = L^0_\pm(\zeta). \label{CI}\ee
 The map $l_0(\zeta)$ is referred to as the {\it extended} solution associated to the ordinary solution $g_0$. Note  also,  that the ordinary
 solution can be extracted from the extended solution since it   coincides with 
 $l_0(0)$  (possibly up to  inessential left multiplication by a constant element from  $G$).
 
 In  the same token,    let $g_\e:W\to G$ be an ordinary solution of the Yang-Baxter $\sigma$-model \eqref{YB} and consider the  Lax pair \eqref{dLaxJ}  of this model obtained in \cite{K09}:
  \be L^{\e}_\pm(\lm) = \biggl(\e^2 \mp \e R -\frac{1+\e^2}{1\pm \lm}\biggr) (I \pm \e R)^{-1}g^{-1}\partial_{\pm}g.\label{dLaxJe}\ee
  (Note that we have introduced the coupling constant $\e$ instead of $\bt$;  the reason for this notation will become clear later).  Because $g_\e$ is solution, the zero curvature  condition holds for generic $\lm$:
    \be \d_+L_-^\e(\lm)-\d_-L_+^\e(\lm)+  [L_-^\e(\lm),L_+^\e(\lm)]=0 \label{ccoe}\ee
    hence there exists a map $l_\e(\lm)$ from   the (simply connected)  world-sheet $W$
  to the complexified group $G^\bc$ such that
 \be - l_\e^{-1} (\lm)\xpm l_\e (\lm) = L^\e_\pm(\lm). \label{Cee}\ee
 The map $l_\e(\lm)$ will be referred to as the {\it extended} solution of the Yang-Baxter $\sigma$-model  associated to the ordinary solution $g_\e$. Note  also,  that the ordinary
 solution $g_\e$  can be extracted from the extended solution since it   coincides with 
 $l_\e(\infty)$.
 
 In what follows, we shall  also need the  concept of the Iwasawa decompositon of the complexified group $G^{\mathbb C}$ \cite{Zhel} in order to decompose in the Iwasawa way the extended solutions just discussed.
  To  explain what is the Iwasawa decomposition, we  first denote   $\G^\bc$ the complexification of $\G$  and view it  as the real Lie algebra.
Clearly, the multiplication by the imaginary unit $\ri $ is $\br$-linear operator from $\G^\bc\to\G^\bc$, so its restriction on $\G$ is well-defined $\br$-linear operator with the domain $\G$ and  the range $\G^\bc$. Thus
 $(R-\ri)$ can be also understood as  a  $\br$-linear operator from $\G $ to $\G^\bc$. Using the modified Yang-Baxter equation \eqref{myb}, it can be easily verified  that  $(R- \ri)$  is in fact 
an injective  homomorpism between the real Lie algebras $\G_R$ and $\G^\bc$ and it thus permits to view $\G_R$ as the real subalgebra of $\G^\bc$. 
 The subgroup $G_R$ of $G^\bc$, obtained  by  integrating
the Lie subalgebra $\G_R$ of $\G^\bc$,  turns out to be nothing but the so called group $AN$.
Recall, that  an element $b$ of  $AN$ can be uniquely represented by means
of the exponential map as follows
$$ b={\rm e}^{\phi}{\rm exp}[\Sigma_{\al>0}v_\al E^\al]\equiv
 {\rm e}^{\phi}n.$$
Here $\al$'s denote the roots of $\G^\bc$, $v_\al$ are complex numbers,
 $E^\al$ are the
step generators
and $\phi$ is an Hermitian element 
 of the Cartan subalgebra
of $\G^\bc$.  In particular, if 
   $G^\bc=SL(n,{\bf C})$, the group $AN$ can be identified
 with the group of 
upper triangular matrices of determinant $1$
and with positive real numbers on the diagonal.

\subsection{Yang-Baxter  Lax pair from principal chiral  model}

The Iwasawa theorem \cite{Zhel} guarantees 
 the existence and the uniqueness of the map Iw: $G^\bc \to G$ 
   such that,   for every $l\in G^\bc$, 
   the    product 
 Iw$(l)l^{-1}$ belongs to $AN$.  Consider now an extended solution  $l_0(\zeta)$ of the principal chiral model associated to some ordinary solution $g_0$ and  define
 \be g_\e:= {\rm Iw}(l_0(-\ri \e)), \quad b_\e:= l_0(-\ri \e)g_\e^{-1}.\ee
 We have therefore
 \be \frac{g_0^{-1}\d_\pm g_0}{1\mp\ri\e}=l_0(-\ri\e)^{-1}\d_\pm l_0(-\ri \e)=g_\e^{-1} b_\e^{-1}\d_\pm b_\e g_\e +g_\e^{-1}\d_\pm g_\e. \label{edm}\ee
 Because $b_\e^{-1}\d_\pm b_\e \in$Lie$(AN)$, there exists  $K_\pm (\xi_+,\xi_-)\in \G$ such that
 \be b_\e^{-1}\d_\pm b_\e =\e(R-\ri)K_\pm.\ee
 This fact  and Eq.\eqref{edm}  permit to infer that
 \be g_0^{-1}\d_\pm g_0   =(1\mp\ri \e) (\e g_\e^{-1}(R-\ri)K_\pm g_\e+g_\e^{-1}\d_\pm g_\e).\label{adm}\ee
 Note that  the left hand side of Eq.(\ref{adm}) is $\G$-valued  but the right hand side is $\G^\bc$-valued. This is possible only  if 
  the $i\G$-part of the right hand side vanishes:
$$i\e g_\e^{-1}(K_\pm \pm \e RK_\pm \pm \d_\pm g_\e  g_\e^{-1})g_\e=0$$
that is
\be K_\pm =\mp(1\pm\e R)^{-1} \d_\pm g_\e g_\e^{-1}.\label{DD}\ee
By inserting \eqref{DD} back into \eqref{adm}, we obtain
\be g_0^{-1}\d_\pm g_0 =  \frac{1+\e^2}{I\pm\e R_{g_\e}}g_\e^{-1} \d_\pm g_\e,\ee
or  
 \be -\frac{g_0^{-1}\d_\pm g_0}{1\pm\zeta}= - \frac{1+\e^2}{1\pm\zeta}(I\pm\e R_{g_\e})^{-1}g_\e^{-1} \d_\pm g_\e. \label{por} \ee
 Note that  on the left hand side of \eqref{por}  is the Lax pair  \eqref{ZM}  of the  principal chiral model and on the right hand side is the Lax pair \eqref{dLaxD}  of the Yang-Baxter $\sigma$-model
 in the version of \cite{D}. 
 
 We remark, that  we were not aware about this simple method to obtain the Lax pair of the Yang-Baxter $\sigma$-model when we had been writing the paper \cite{K09}. As a matter of fact,
 however, we are  benefiting from this circumstance here. Why? Because the alternative version \eqref{dLaxJ} of the Yang-Baxter Lax pair, that we have obtained in \cite{K09} in a more complicated way, serves in this article as the starting point for obtaining the bi-Yang-Baxter Lax pair by the simple  method presented in this subsection. Surprisingly enough,  the extended solution obtained from the 
other  Lax pair \eqref{dLaxD}  turns out to be of no direct utility in this respect, as it can be easily checked.

 \subsection{Bi-Yang-Baxter Lax pair from Yang-Baxter model}
 
 Consider now the  extended solution  $l_\e(\lm)$ of the  Yang-Baxter $\sigma$-model associated to some ordinary solution $g_\e$. Recall that $l_\e(\lm)$ verifies \eqref{Cee}. For  a real $\eta$,  we define
 \be g_{\e\eta}:= {\rm Iw}(l_\e(-\ri \eta)), \quad b_{\e\eta}= l_\e(-\ri \eta)g_{\e\eta}^{-1}.\ee
 We have therefore
 
 $$   -\biggl(\e^2 \mp \e R -\frac{1+\e^2}{1\mp \ri \eta}\biggr) (I \pm \e R)^{-1}g_\e^{-1}\partial_{\pm}g_\e=
 l_\e(-\ri\eta)^{-1}\d_\pm l_\e(-\ri \eta)=$$\be =g_{\e\eta}^{-1}b_{\e\eta}^{-1}\d_\pm b_{\e\eta} g_{\e\eta}+g_{\e\eta}^{-1}\d_\pm g_{\e\eta}=
 g_{\e\eta}^{-1}\eta(R-\ri)K_\pm g_{\e\eta}+g_{\e\eta}^{-1}\d_\pm g_{\e\eta}. \label{edmb}\ee
 Indeed, because  of $b_{\e\eta}^{-1}\d_\pm b_{\e\eta} \in$Lie$(AN)$, there exists  $K_\pm (\xi_+,\xi_-)\in \G$ such that
 \be b_{\e\eta}^{-1}\d_\pm b_{\e\eta}  =\eta(R-\ri)K_\pm.\ee
In analogy with the previous subsection, we know infer from Eq.\eqref{edmb}   that the following expression must  have the vanishing $i\G$-part :
\be \biggl(\e^2 \mp \e R -\frac{1+\e^2}{1\pm \ri \eta}\biggr)  \biggl(g_{\e\eta}^{-1}\eta(R-\ri)K_\pm g_{\e\eta}+g_{\e\eta}^{-1}\d_\pm g_{\e\eta}\biggr).\label{admb}\ee
This condition permits to determine $K_\pm$:
\be K_\pm =\mp \frac{1+\e^2}{1-\e^2\eta^2}\biggl(I\pm\frac{\eta(1+\e^2)}{1-\e^2\eta^2}R\pm \frac{\e(1+\eta^2)}{1-\e^2\eta^2}R_{g^{-1}_{\e\eta}}\biggl)^{-1} \d_\pm g_{\e\eta}g_{\e\eta}^{-1}.\label{ddd}\ee
 By inserting \eqref{ddd} back into \eqref{edmb}, we obtain after some calculation

  \be (I \pm \e R)^{-1}g_\e^{-1}\partial_{\pm}g_\e=\frac{1+\eta^2}{1-\e^2\eta^2}\biggl(I\pm\frac{\eta(1+\e^2)}{1-\e^2\eta^2}R_{g_{\e\eta}}\pm \frac{\e(1+\eta^2)}{1-\e^2\eta^2}R\biggr)^{-1}g_{\e\eta}^{-1}\partial_\pm g_{\e\eta} \ee
  hence
  $$ \biggl(\e^2 \mp \e R -\frac{1+\e^2}{1\pm \lm}\biggr) (I \pm \e R)^{-1}g_\e^{-1}\partial_{\pm}g_\e=$$ $$=  \biggl(\e^2 \mp \e R -\frac{1+\e^2}{1\pm \lm}\biggr) \frac{1+\eta^2}{1-\e^2\eta^2}\biggl(I\pm\frac{\eta(1+\e^2)}{1-\e^2\eta^2}R_{g_{\e\eta}}\pm \frac{\e(1+\eta^2)}{1-\e^2\eta^2}R\biggr)^{-1}g_{\e\eta}^{-1}\partial_\pm g_{\e\eta} =$$
  \be  = \mp \biggl(\beta(R-\ri)+ \frac{2\ri \bt \pm(1+\al^2-\bt^2)}{1\pm\zeta}\biggr) (I\pm\al R_{g_{\e\eta}}\pm \bt R)^{-1}g_{\e\eta}^{-1}\partial_{\pm}g_{\e\eta},\label{final}\ee
  where \be \zeta=\frac{\lm+\ri\e}{1+\ri\e\lm},\qquad  \al\equiv \frac{\eta(1+\e^2)}{1-\e^2\eta^2}, \qquad \bt \equiv  \frac{\e(1+\eta^2)}{1-\e^2\eta^2}.\ee
  Note that  on the extreme left hand side of \eqref{final}  is the Lax pair  \eqref{dLaxJ}  of the  Yang-Baxter $\sigma$-model and on the  extreme right hand side is the Lax pair \eqref{dLax}  of the bi-Yang-Baxter $\sigma$-model.

 \medskip

   \section{Conclusions and outlook}

The principal result of this paper  is the
   construction of the Lax pair (\ref{dLax}) of the bi-Yang-Baxter $\si$-model.   As far as the directions to develop further our work,  we think that a promising one should
consist in the study of the  
  T-duality story which is naturally associated to any Poisson-Lie
   symmetric $\si$-model \cite{KS95a}. The bi-Yang-Baxter model  is doubly Poisson-Lie symmetric and hence  Poisson-Lie T-dualizable from both right and left side.
Combining the left and the  right T-duality, a novel nontrivial dynamical
equivalence
of two seemingly different $\si$-models living on the target of the   dual Poisson-Lie group $AN$ should be  thus obtained.
 Another natural open  problem  is to find out whether the bi-Yang-Baxter Lax pair  constructed  in the present work can be obtained by a suitable adaptation of the method of  Ref.\cite{DMV13}.

  \end{document}